%% file: main.tex
\documentclass[sigconf]{acmart}
\AtBeginDocument{%
  \providecommand\BibTeX{{%
    \normalfont B\kern-0.5em{\scshape i\kern-0.25em b}\kern-0.8em\TeX}}}
\usepackage{amsmath,amsfonts}
\usepackage{algorithmic}
\usepackage{graphicx}
\usepackage[caption=false]{subfig}
\usepackage{float}
\usepackage{multirow}
\usepackage{array}
\graphicspath{{figs/}}
\usepackage{url}
\usepackage{caption}
\copyrightyear{2023} 
\acmYear{2023} 
\setcopyright{rightsretained} 
\acmConference[SNTA '23]{Proceedings of the 2023 Systems and Network Telemetry and Analytics}{June 20, 2023}{Orlando, FL, USA}
\acmBooktitle{Proceedings of the 2023 Systems and Network Telemetry and Analytics (SNTA '23), June 20, 2023, Orlando, FL, USA}
\acmDOI{10.1145/3589012.3594897}
\acmISBN{979-8-4007-0165-8/23/06}

\usepackage{etoolbox}
\makeatletter
\patchcmd{\maketitle}{\@copyrightpermission}{
\begin{minipage}{0.3\columnwidth}
\href{http://creativecommons.org/licenses/by/4.0/}{\includegraphics[width=0.90\textwidth]{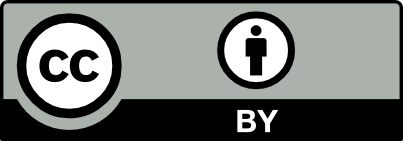}}
\end{minipage}\hfill
\begin{minipage}{0.7\columnwidth}
\href{http://creativecommons.org/licenses/by/4.0/}{This work is licensed under a Creative Commons Attribution International 4.0 License.}
\end{minipage}

\vspace{5pt}
}{}{}

\makeatother

\begin{document}
\fancyhead{}
\title{Analyzing Transatlantic Network Traffic over Scientific Data Caches}

\author{Ziyue Deng}
\affiliation{%
  \institution{University of California, Berkeley}
  \city{Berkeley}
  \state{CA}
  \country{USA}
}
\email{ziyue_deng@berkeley.edu}

\author{Alex Sim}
\affiliation{%
  \institution{Lawrence Berkeley National Laboratory}
  \city{Berkeley}
  \state{CA}
  \country{USA}
}
\email{asim@lbl.gov}

\author{Kesheng Wu}
\affiliation{%
  \institution{Lawrence Berkeley National Laboratory}
  \city{Berkeley}
  \state{CA}
  \country{USA}
}
\email{kwu@lbl.gov}

\author{Chin Guok}
\affiliation{%
  \institution{Energy Sciences Network}
  \city{Berkeley}
  \state{CA}
  \country{USA}
}
\email{chin@es.net}

\author{Damian Hazen}
\affiliation{%
  \institution{Energy Sciences Network}
  \city{Berkeley}
  \state{CA}
  \country{USA}
}
\email{dhazen@es.net}

\author{Inder Monga}
\affiliation{%
  \institution{Energy Sciences Network}
  \city{Berkeley}
  \state{CA}
  \country{USA}
}
\email{imonga@es.net}

\author{Fabio Andrijauskas}
\affiliation{%
  \institution{University of California, San Diego}
  \city{La Jolla}
  \state{CA}
  \country{USA}
}
\email{fandrijauskas@sdsc.edu}

\author{Frank W\"{u}rthwein}
\affiliation{%
  \institution{University of California, San Diego}
  \city{La Jolla}
  \state{CA}
  \country{USA}
}
\email{fkw@ucsd.edu}

\author{Derek Weitzel}
\affiliation{%
  \institution{University of Nebraska, Lincoln}
  \city{Lincoln}
  \state{NE}
  \country{USA}
}
\email{dweitzel@unl.edu}

\renewcommand{\shortauthors}{Deng, et al.}

\begin{abstract}
Large scientific collaborations often share huge volumes of data around the world.
Consequently a significant amount of network bandwidth is needed for data replication and data access.
Users in the same region may possibly share  resources as well as data, especially when they are working on related topics with similar datasets.
In this work, we study the network traffic patterns and resource utilization for scientific data caches connecting European networks to the US. 
We explore the efficiency of  resource utilization, especially for network traffic which consists mostly of transatlantic data transfers, and the potential for having more caching node deployments. 
Our study shows that these data caches reduced  network traffic volume by 97\% during the study period. 
This demonstrates that such caching nodes are effective in reducing wide-area network traffic. 
\end{abstract}

\ccsdesc[500]{Networks~Network performance analysis}
\ccsdesc[500]{Computing methodologies~Distributed computing methodologies}

\keywords{network cache, osdf, resource utilization, data pattern, xcache}

\maketitle

\input{intro}
\input{background}
\input{data}
\input{eval}
\input{conc}

\begin{acks}
This work was supported by the Office of Advanced Scientific Computing Research, Office of Science, of the U.S. Department of Energy under Contract No. DE-AC02-05CH11231, and also used resources of the National Energy Research Scientific Computing Center (NERSC). 
This work was also supported by the National Science Foundation through the grants OAC-2030508, OAC-1836650, MPS-1148698, PHY-1120138, and OAC-1541349. This work is also supported by the US CMS M\&O 2121686, PNRP: NSF 2112167, and PRP: NSF OAC-1541349 
\end{acks}

\bibliographystyle{ACM-Reference-Format}
\bibliography{main}

\end{document}

%% file: intro.tex
\section{Introduction}
\label{sec:intro}

Scientific experiments and simulations generate large volumes of data over time.
Such data is shared by geographically distributed users, which creates a large amount of network traffic for data replication and access during analysis~\cite{esnetHepReq}. 
Data storage caches have been deployed for regional users engaged in related study topics~\cite{socalrepo2018, datalakes, osg, stashcache, xcache2014}. 
These storage caches can hold a significant portion of the datasets close to user accesses, which reduces the data access latency and improves data analysis throughput~\cite{copps2021, han2022, sim2023}. 

One set of such storage caches is deployed by the Open Science Data Federation (OSDF). 
This study explores the efficiency of resource utilization of these OSDF caches for transatlantic data transfers.
More specifically, we focus on two caching nodes for data transfers from US to Europe. 
Understanding their resource utilization could lead to more efficient management of future cache deployments.

The contributions of this paper can be summarized as follows: 
(1) our study finds that the data caches reduce the network traffic volume by 97\% during the study period; 
(2) this network traffic reduction is from transatlantic traffic from the US to Europe, which is what the data cache nodes are designed for;
(3) based on the observed network traffic reduction rates, we can plan for additional deployments of the caching nodes to benefit science communities.

%% file: background.tex
\section{Background}
\label{sec:background}



\begin{figure}[ht]
         \centering
         \includegraphics[width=\linewidth, height=4.0cm]  {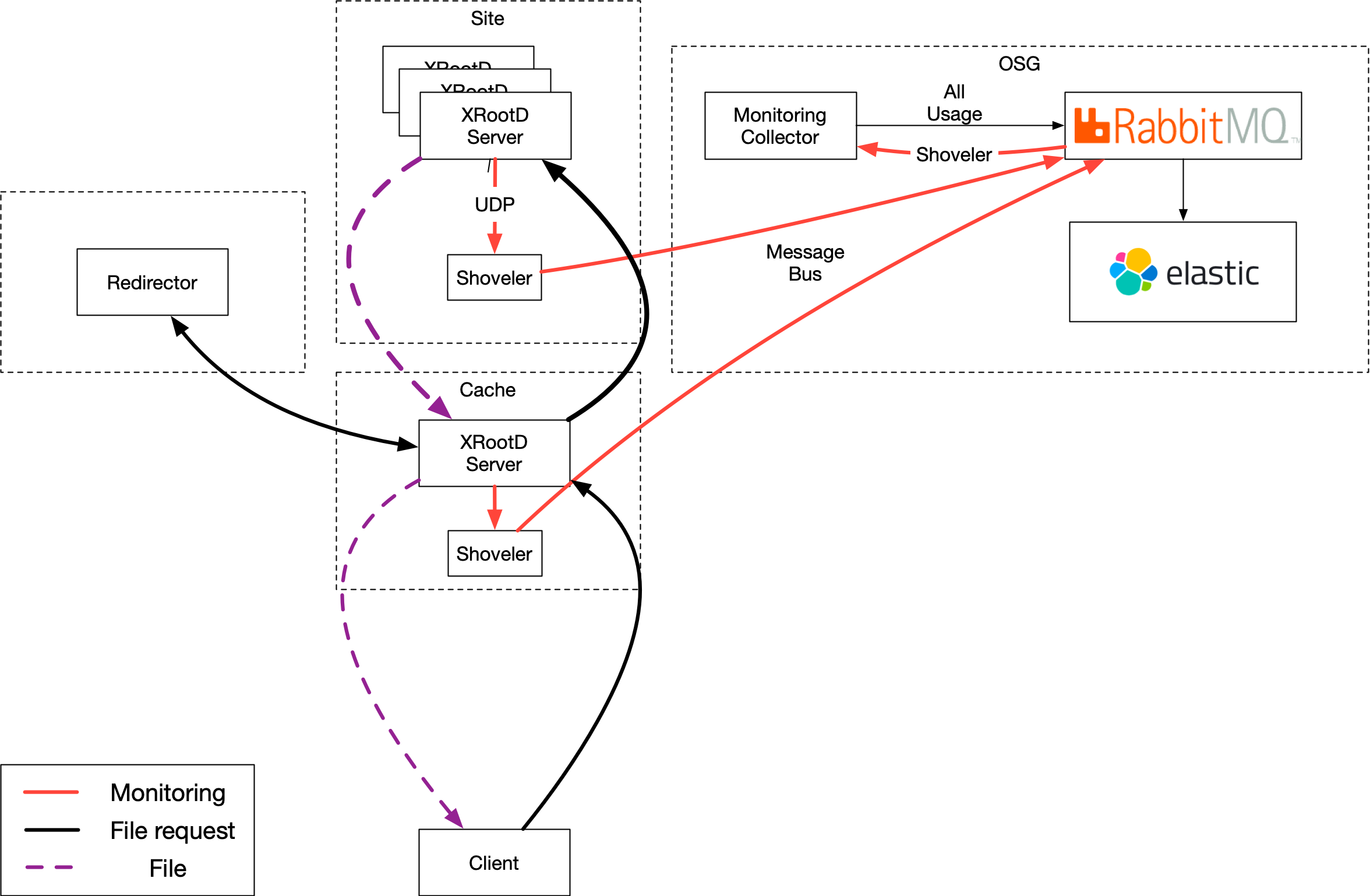}  
         \caption{Open Science Data Federation.}
         \label{fig:osg}
\vspace{-0.2cm}
\end{figure}

The Open Science Grid (OSG) provides data access resources for High Throughput Computing (HTC) facilities~\cite{10.1145/3332186.3332212}. 
The OSG data infrastructure is named the Open Science Data Federation (OSDF). 
It holds data files from several large experiments and many smaller independent projects. 
It supports massive amounts of data from National Science Foundation (NSF) funded projects. 
At the core of the OSDF are the concepts of "data origin," "data caches," and "data access redirector," all implemented as services via XrootD~\cite{xrootd2005}.  
Figure~\ref{fig:osg} shows simplified schematics of these components. 

Origins host the data.  Multiple origins are tied into a tree structure to form the data federation. 
Applications access the OSDF generally via the closest cache to the compute host they execute on. 
The meaning of closest is determined via GeoIP~\cite{weitzel2019stashcache}. 
All services are deployed as containers via Kubernetes. 
When a job requests a file from OSDF, usually this file request is satisfied with a replica from the cache. 
However, if the file is unavailable in the cache, the cache asks the redirector to locate a copy. 
This file is copied from the origin to the cache, and the cache sends it to the client. On all OSDF levels, file accesses are monitored.

XrootD has several streams that provide monitoring information, such as cache monitoring, transfer information, and file access monitoring.
Caches and Origins use ``shoveler'' software to send the monitoring streams to the queue message system.  The collector processes these messages and sends them back to the queue system.
In the end, all the monitoring data goes to a central logging system managed by RabbitMQ, as shown in Figure~\ref{fig:osg}.

This study examines logs from one cache node in Cardiff, UK, and another in Amsterdam, Netherlands. These two cache nodes are mainly designed to serve those data from the US to Europe. On 10 Gps networks, Cardiff and Amsterdam nodes have 81 TB and 30 TB of storage capacities respectively. 

%% file: data.tex
\section{Datasets}
\label{sec:data}

Our work is based on the logs collected from the Amsterdam and Cardiff nodes between October 2022 and January 2023.
The collected information includes the following attributes about every data access request: timestamp, access count, file path, project group (VO), file size, attach time, detach time, remote origin, bytes hit cache, bytes miss cache, and bytes bypass cache. 
A total of 399,321 data access requests from the Cardiff node and 31,508,228 data access requests from the Amsterdam node are included in this study.

\begin{table}
\scriptsize
\centering
\caption{
Summary statistics for data accesses from the Cardiff node}

\begin{tabular}{|c||c|c|c|c|c|} \hline
  & \shortstack{\# of \\Requests} & \shortstack{\# of \\Cache Miss\\Requests} & \shortstack{Total Size\\of Requested\\Bytes (GB)} & \shortstack{Total Size\\ of Cache Misses\\(GB)} & \shortstack{\% of Cache\\Miss Size} \tabularnewline \hline \hline
noVO &  10,807 & 131 & 0.00698 & 0.000134 & 1.92\%  \tabularnewline \hline
osg  & 388,484 & 96,083 & 38,454 & 1,173.37 & 3.05\% 
\tabularnewline \hline
gwdata & 30 & 5 & 7.78 & 0.109 & 1.40\%  \tabularnewline \hline \hline
Total &  399,321 & 96,219 & 38,461 & 1,173.48 & 3.05\%  \tabularnewline \hline
\end{tabular}
\vspace{-0.2cm}
\label{tab:summary_data_cardiff}
\end{table}

Table~\ref{tab:summary_data_cardiff} shows basic statistics about the data accesses for the Cardiff node for each Virtual Organization (VO). 
A total of 399K data requests were received, and about 24\% of the total requests are cache misses.
When the requested file is not in the cache, it is retrieved from the remote origins which are mostly from the US. 
The total requested data volume was 38.4 TB, among which 1.17 TB was transferred over the wide-area network. 
The percentage of cache misses is calculated by (Total Size of Cache Misses) / (Total Size of Requested Bytes).
For the Cardiff node, the percentage of cache misses is very low; less than 4\%.
The VO osg has the highest percentage of cache misses at 3.05\%, and the gwdata (LIGO) project has the lowest percentage of cache misses at 1.40\%.

\begin{table}
\scriptsize
\centering
\caption{
Monthly summary statistics for data accesses from the Amsterdam node}

\begin{tabular}{|c||c|c|c|c|c|} \hline
  & \shortstack{\# of \\Requests} & \shortstack{\# of \\Cache Miss\\Requests} & \shortstack{Total Size\\of Requested\\Bytes (GB)} & \shortstack{Total Size\\ of Cache Misses\\(GB)} & \shortstack{\% of Cache\\Miss Size} \tabularnewline \hline \hline
Oct 2022 & 9,290,400 & 2,931 & 294,017 & 79 & 0.03\%  \tabularnewline \hline
Nov 2022 & 13,160,620 & 19,729& 473,367 & 348 & 0.07\%  
\tabularnewline \hline
Dec 2022 & 6,140,170 & 26,765 & 140,843 & 327 & 0.23\%
\tabularnewline \hline
Jan 2023 & 2,917,038 & 13,680& 83,343 & 617 & 0.74\%  
\tabularnewline \hline \hline
Total &  31,508,228 & 63,105 & 991,570 & 1,371 & 0.14\%  \tabularnewline \hline
\end{tabular}
\vspace{-0.4cm}
\label{tab:summary_data_Ams_month}
\end{table}

\begin{table}
\scriptsize
\centering
\caption{
Summary statistics for data accesses per VO from the Amsterdam node}

\begin{tabular}{|c||c|c|c|c|c|} \hline
  & \shortstack{\# of \\Requests} & \shortstack{\# of \\Cache Miss\\Requests} & \shortstack{Total Size\\of Requested\\Bytes (GB)} & \shortstack{Total Size\\ of Cache Misses\\(GB)} & \shortstack{\% of Cache\\Miss Size} \tabularnewline \hline \hline
fermilab & 6,854,234 & 6,401 & 109,840 & 104 & 0.10\%  
\tabularnewline \hline
osg & 24,493,597 & 39,125 & 877,349 & 928 & 0.11\%  
\tabularnewline \hline
noVO & 78,502 & 10,386 & 0.0789 & 0.0106 & 13.48\% 
\tabularnewline \hline
gwdata & 72,865 & 5,474 & 1,825 & 169 & 9.25\% 
\tabularnewline \hline
icecube & 9,030 & 1,719 & 2,556 & 169 & 6.61\% 
\tabularnewline \hline
\end{tabular}
\vspace{-0.4cm}
\label{tab:summary_data_Ams_vo}
\end{table}

\begin{table}
\scriptsize
\centering
\caption{
Summary statistics for data accesses for DUNE and LIGO projects from the Amsterdam node}

\begin{tabular}{|c||c|c|c|c|c|} \hline
  & \shortstack{\# of \\Requests} & \shortstack{\# of \\Cache Miss\\Requests} & \shortstack{Total Size\\of Requested\\Bytes (GB)} & \shortstack{Total Size\\ of Cache Misses\\(GB)} & \shortstack{\% of Cache\\Miss Size} \tabularnewline \hline \hline
DUNE & 5,006,143 & 102 & 99929 & 0.24     & 0.00024\%  
\tabularnewline \hline
LIGO & 1,981,630  & 38,593  & 164,451  &  927 & 0.56\%  
\tabularnewline \hline
\end{tabular}
\vspace{-0.3cm}
\label{tab:summary_data_Ams_projects}
\end{table}

Table~\ref{tab:summary_data_Ams_month} and Table~\ref{tab:summary_data_Ams_vo} show basic statistics about data accesses to the Amsterdam node during the study period, grouped by month and VO respectively.
Table~\ref{tab:summary_data_Ams_month} shows the highest data access requests in November 2022 with about 13 million requests, but the percentage of cache miss sizes is low at 0.07\%.
Cache misses by volume ranging from 0.03\% to 0.74\% is observed to be low across all months; less than 1\%, which means that most of the requested data are already in the cache.

Table~\ref{tab:summary_data_Ams_vo} shows that two VOs have a high number of requests with low cache misses by volume.
This indicates that users from these two VOs have similar research interests and are accessing the same dataset. 
On the other hand, the other two VOs have a relatively lower number of data access requests with higher cache misses.

Table~\ref{tab:summary_data_Ams_projects} shows summary statistics about data accesses for the Deep Underground Neutrino Experiment (DUNE) and the Laser Interferometer Gravitational-Wave Observatory (LIGO) projects on Amsterdam node.  
DUNE and LIGO projects are identified from the data paths.
The two projects have a high number of requests with very low number and volume of cache misses.
This indicates that users from these two projects tend access the same set of files.

%% file: eval.tex
\section{Resource Utilization}
\label{sec:eval}
Next, we study the effectiveness of the storage caches of OSDF by exploring two types of statistics: daily file requests and network traffic avoided by the use of these caches.

\subsection{Cache Utilization}

\begin{figure}
\centering
\vspace{-0.4cm}
\subfloat[Cardiff node]{%
  \includegraphics[width=\linewidth, height=3.2cm]{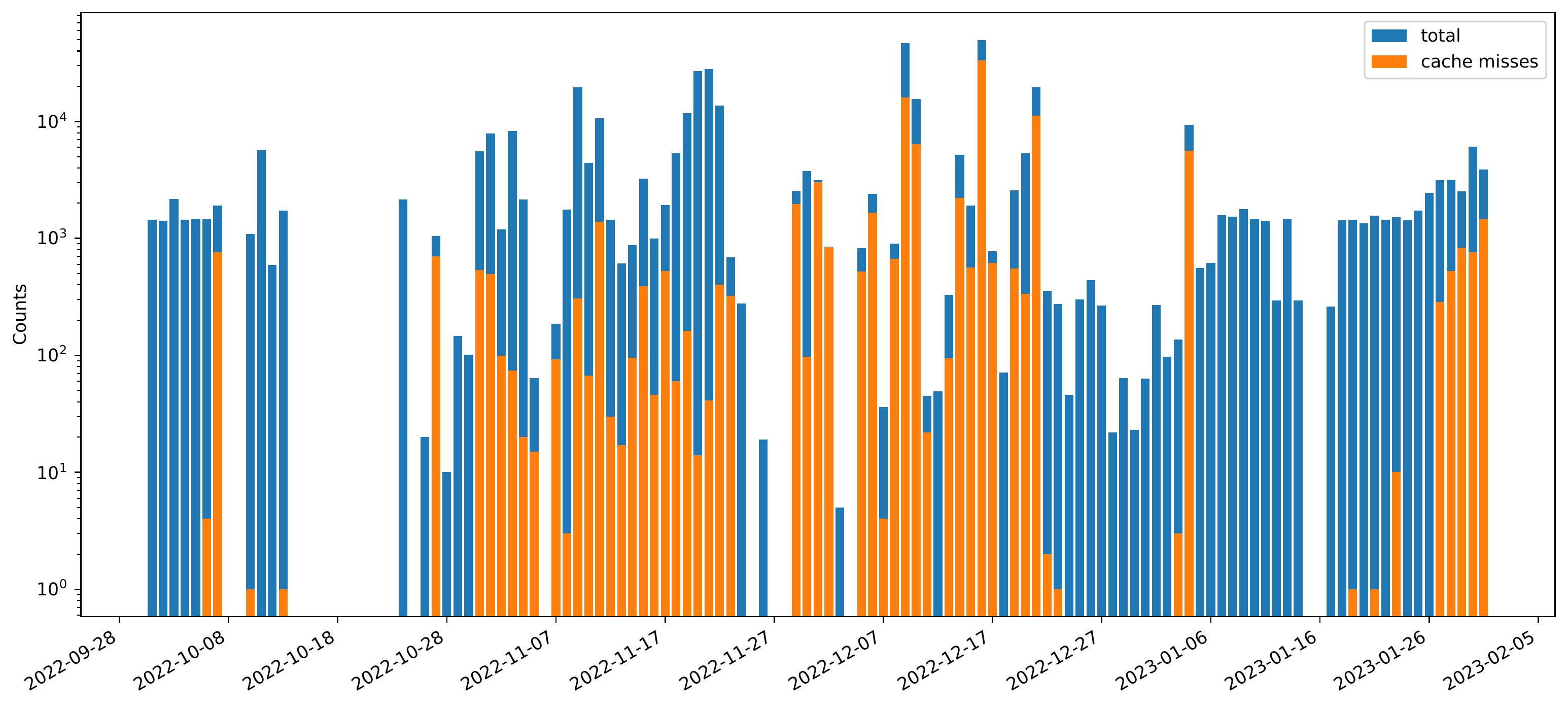}
  \label{fig:cardiff_daily_count}
} \newline
\vspace{-0.2cm}
\subfloat[Amsterdam node]{
  \includegraphics[width=\linewidth, height=3.2cm]{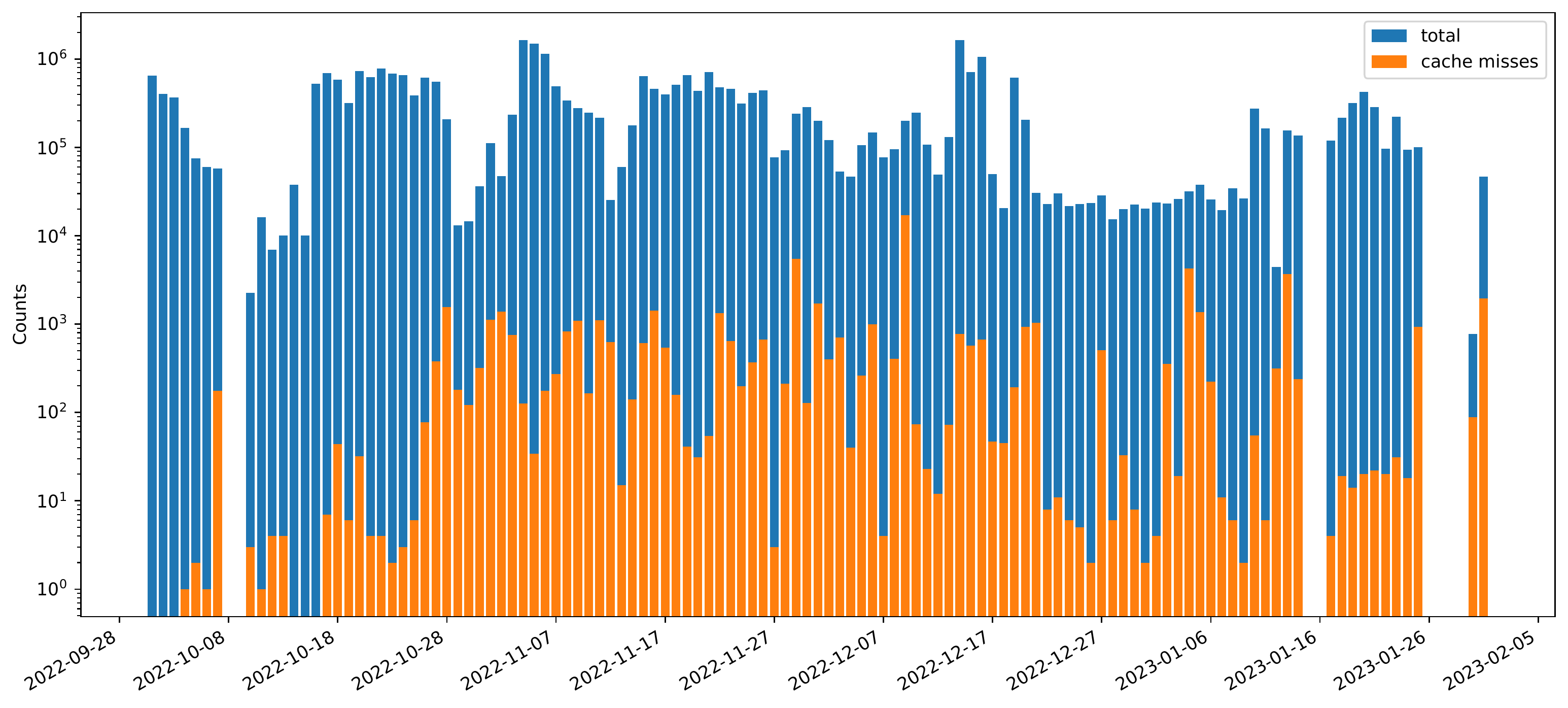}
  \label{fig:ams_daily_count}
}
\caption{Daily number of data requests (in blue) and cache misses (in orange) in log sale from Oct.~2022 to Jan.~2023.  Note that where the number of cache misses are typically orders of magnitude fewer than the number of files requested there are a few days where most of the file requests are cache misses on the cache at Cardiff.}
\label{fig:daily_count_by_site}
\vspace{-0.3cm}
\end{figure}

Figure~\ref{fig:cardiff_daily_count} and Figure~\ref{fig:ams_daily_count} show the daily number of data requests (in blue) and number of cache misses (in orange) in log scale during the study period for the Cardiff node and Amsterdam node respectively. 
The daily number of data accesses for the Cardiff node has large variance throughout the study period, and the overall cache miss rate is 24.1\%.
For the Amsterdam node, the daily number of data accesses has some variance throughout the study period with low cache miss rates. Overall cache miss rate for  data access requests on the Amsterdam node is 0.2\%. 

\begin{figure}
\centering
\subfloat[DUNE project]{%
  \includegraphics[width=\linewidth, height=3.2cm]{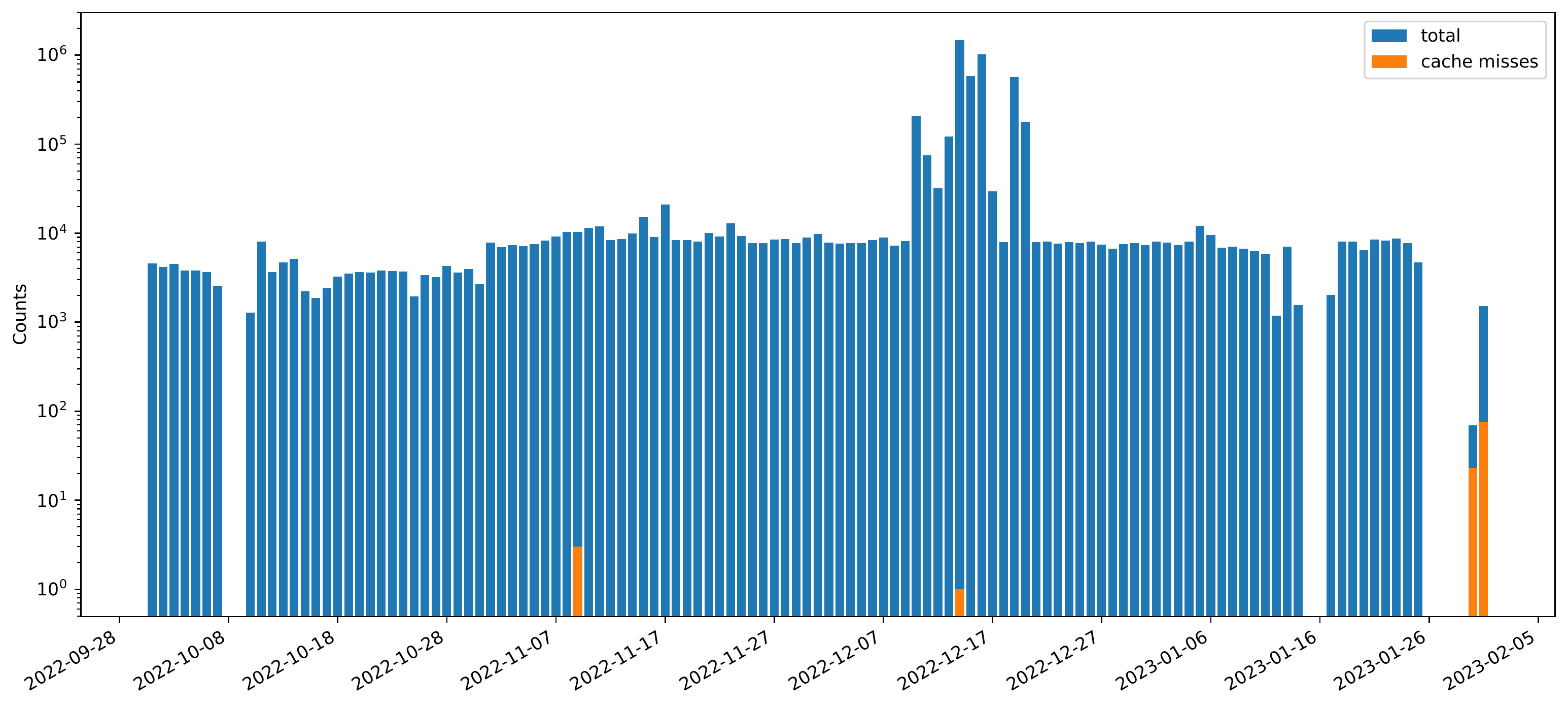}
  \label{fig:dune_daily_count}
} \newline
\vspace{-0.2cm}
\subfloat[LIGO project]{
  \includegraphics[width=\linewidth, height=3.2cm]{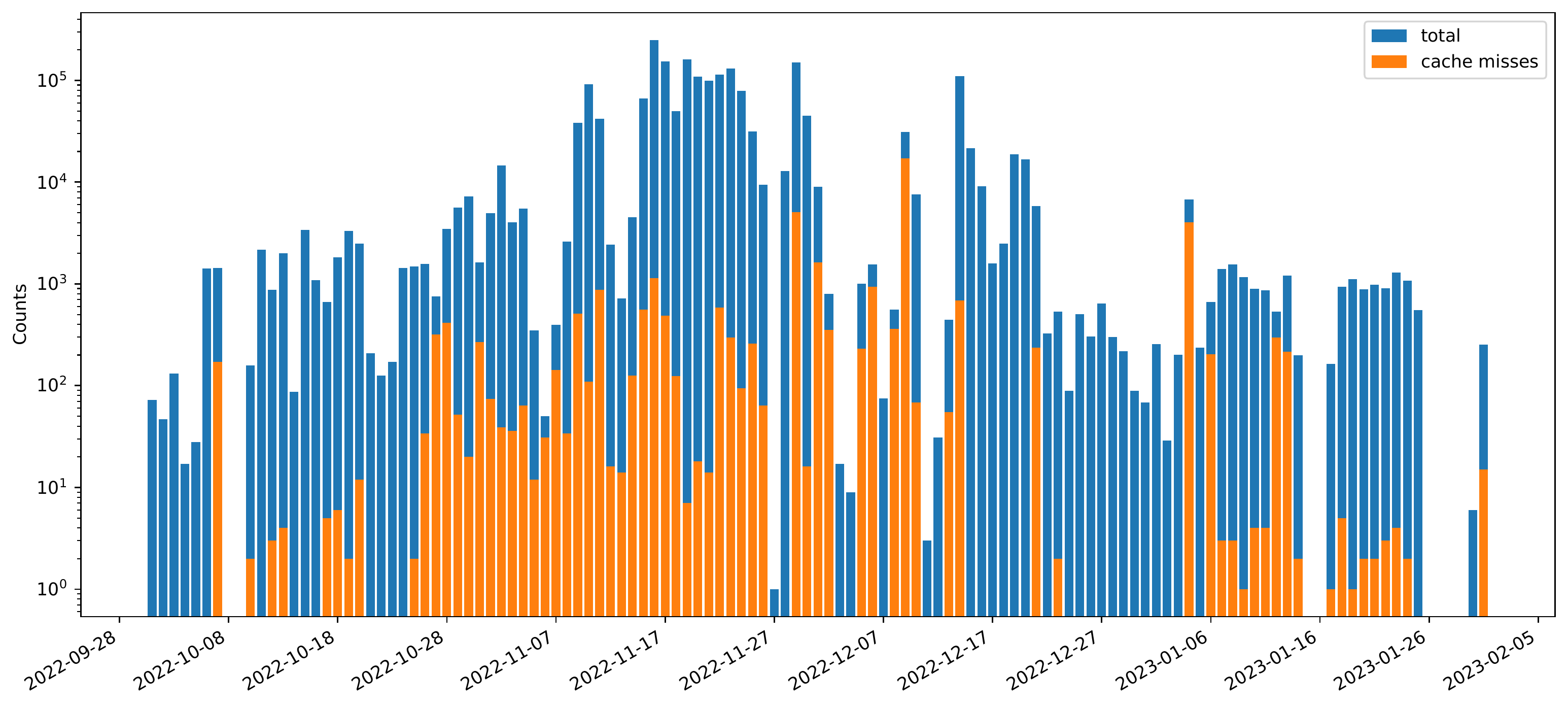}
  \label{fig:ligo_daily_count}
}
\caption{Daily number of data requests (in blue) and cache misses (in orange) in log sale for Amsterdam node from Oct.~2022 to Jan.~2023.  Note there are a couple of days where most of the accesses to LIGO files are cache misses.}
\vspace{-0.4cm}
\label{fig:daily_count_by_vo}
\end{figure}

Figure~\ref{fig:dune_daily_count} and Figure~\ref{fig:ligo_daily_count} show the daily number of data requests (in blue) and number of cache misses (in orange) in log scale on the Amsterdam node for DUNE and LIGO projects respectively. 
The daily number of data accesses is fairly consistent throughout the study period for the DUNE project, and it shows a very low number of cache misses during the study period.
For the LIGO project, the number of data accesses has a large variance throughout the study period, and it has a higher number of cache misses than DUNE project.

\subsection{Network Traffic Reduction}
The file requests served from the cache nodes are reducing  network traffic on the wide-area network.
Next, we examine the cache hits and misses based on the volume of data involved.

\begin{figure}
\centering
\subfloat[Cardiff node]{%
      \includegraphics[width=\linewidth, height=3.2cm]{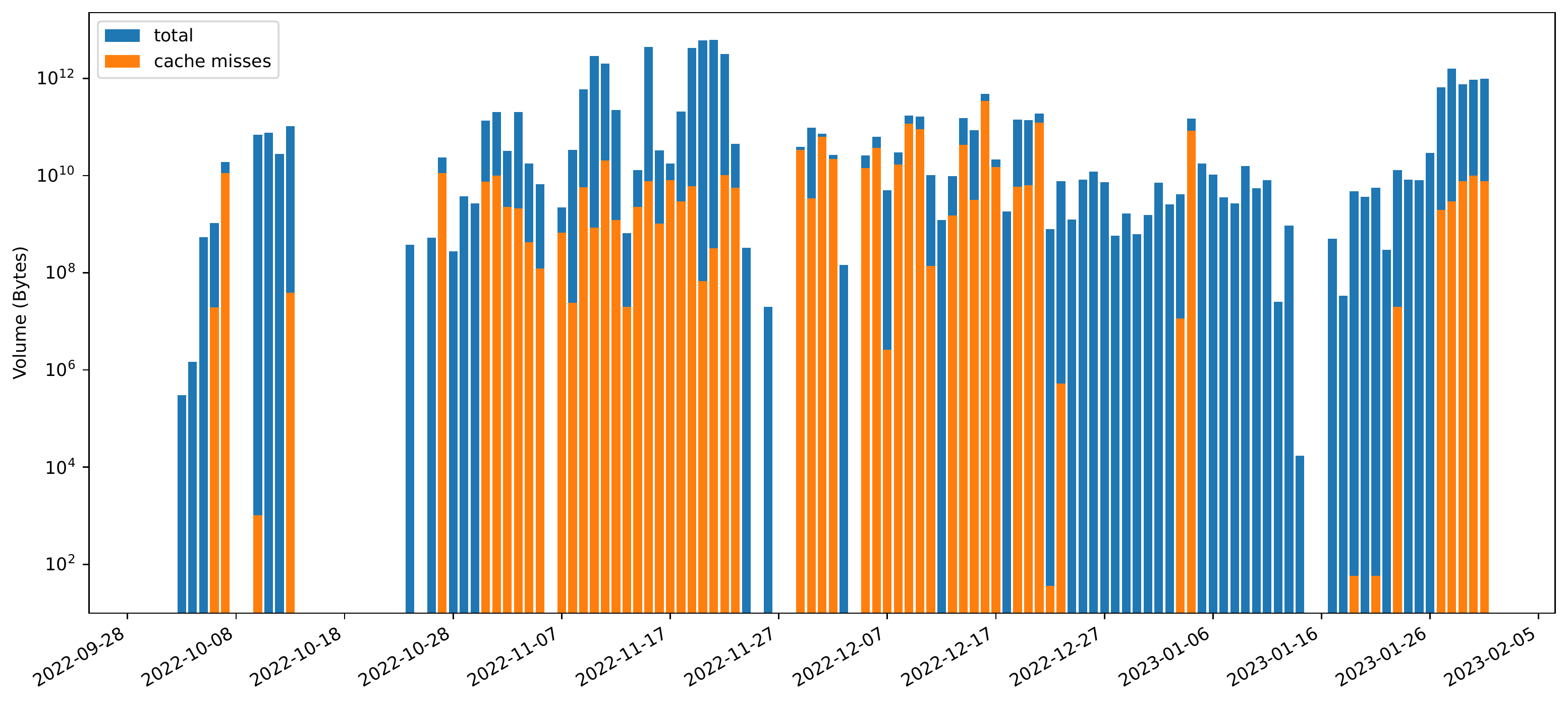}
      \label{fig:Cardiff_Daily_Byte}
} \newline
\vspace{-0.2cm}
\subfloat[Amsterdam node]{
      \includegraphics[width=\linewidth, height=3.2cm]{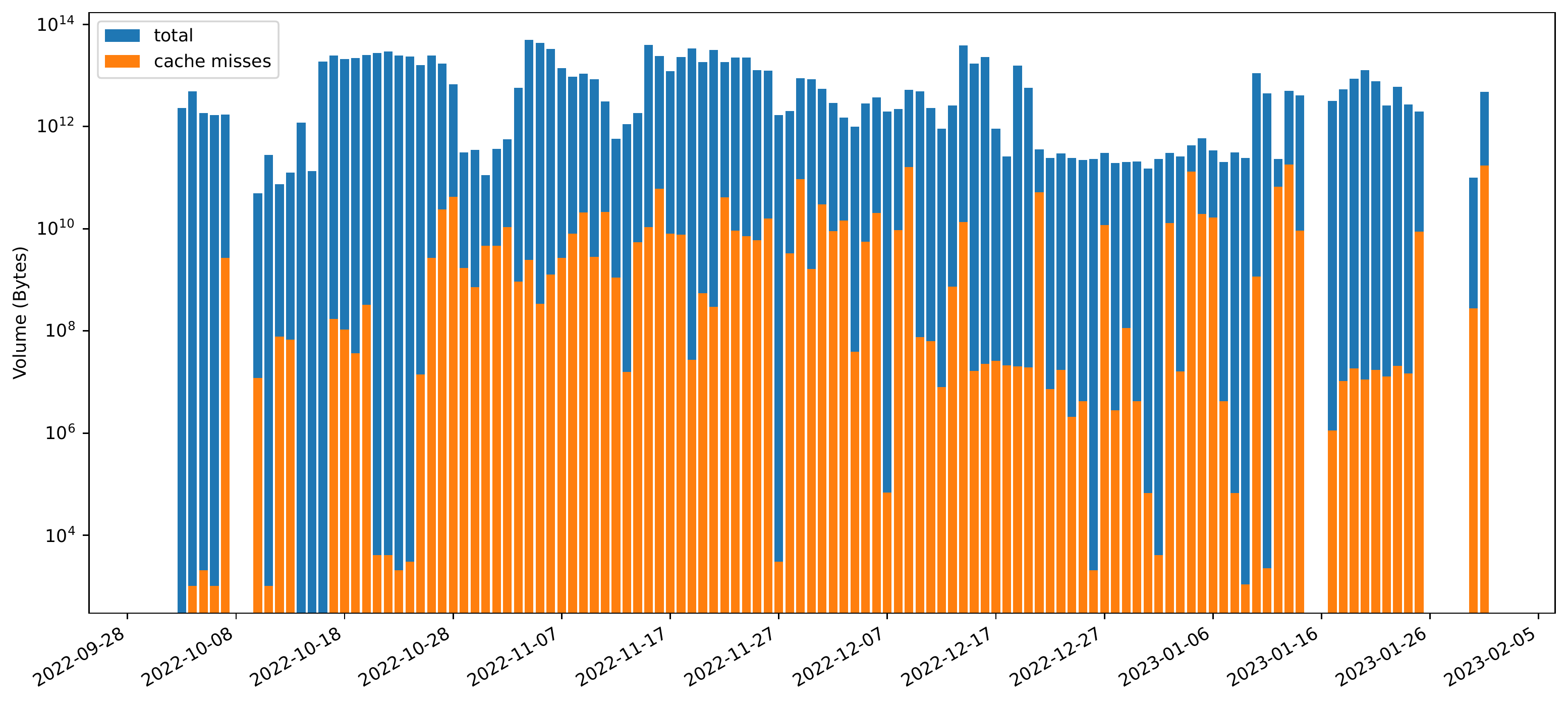}
      \label{fig:ams_Daily_Byte}
}
\caption{Daily volume of requested data (in blue) and cache misses (in orange) in log scale from Oct.~2022 to Jan.~2023.}
\vspace{-0.4cm}
\label{fig:volume_by_site}
\end{figure}

Figure~\ref{fig:Cardiff_Daily_Byte} and Figure~\ref{fig:ams_Daily_Byte} show the daily volume of requested data (in blue) and volume of cache misses (in orange) in log scale for the Cardiff node and Amsterdam node respectively.
Daily requested data volumes on the Amsterdam node in Figure~\ref{fig:ams_Daily_Byte} follows a roughly similar pattern to the daily number of data accesses in Figure~\ref{fig:ams_daily_count}. 
Daily volumes of cache misses on the Amsterdam node in Figure~\ref{fig:ams_Daily_Byte} also follow a similar pattern proportionally to the daily number of cache misses in Figure~\ref{fig:ams_daily_count}, as the figures are in log scale. 
Overall cache miss rate for the data volume on the Amsterdam node is 0.14\%, and the cache miss rate for data access requests is 0.2\%.
It indicates that the average data size for each cache miss is slightly smaller than the average data size for each data request during the study period. 
From the Table~\ref{tab:summary_data_Ams_month}, the average file size for each cache miss is 21.7 MB, and the average data size for each data request is 31.4 MB. 


\begin{figure}
\centering
\subfloat[DUNE project]{%
      \includegraphics[width=\linewidth, height=3.2cm]{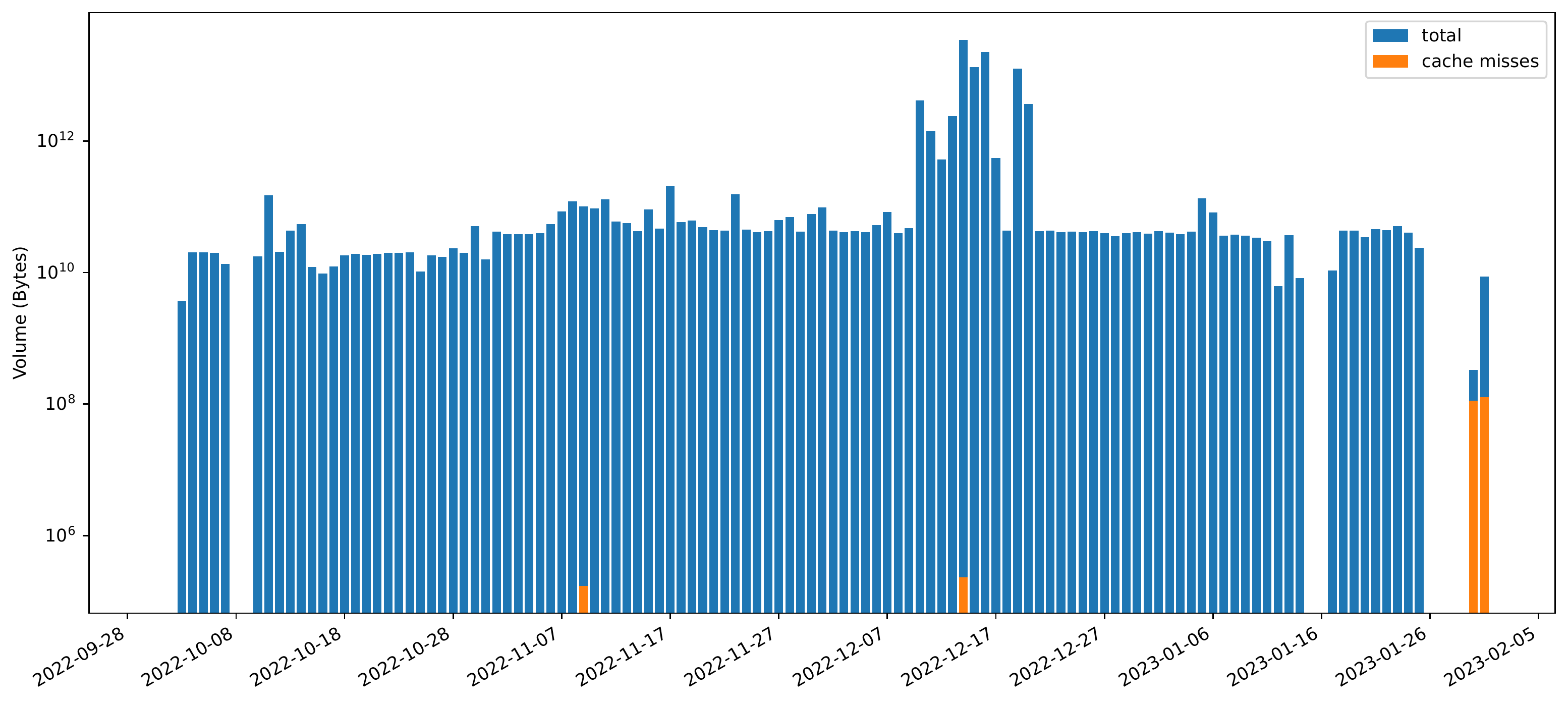}
      \label{fig:dune_Daily_Byte}
} \newline
\vspace{-0.2cm}
\subfloat[LIGO project]{
      \includegraphics[width=\linewidth, height=3.2cm]{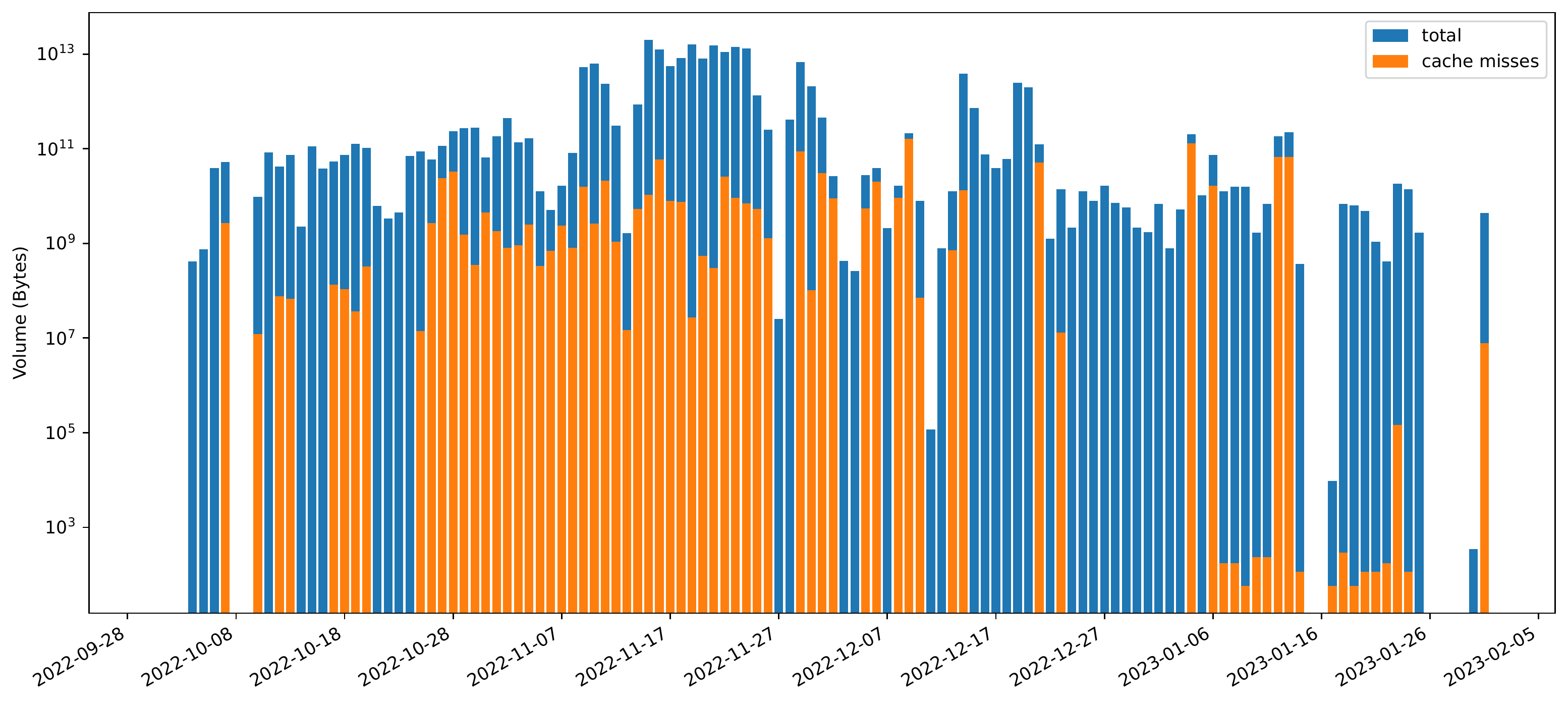}
      \label{fig:ligo_Daily_Byte}
}
\caption{Daily volume of requested data (in blue) and volume of cache misses (in orange) in log scale for Amsterdam node from Oct.~2022 to Jan.~2023}
\vspace{-0.4cm}
\end{figure}

Figure~\ref{fig:dune_Daily_Byte} and Figure~\ref{fig:ligo_Daily_Byte} show the daily volume of requested data (in blue) and volume of cache misses (in orange) in log scale for the DUNE and LIGO projects respectively.
The daily volume of requested data and cache misses for the DUNE project  in Figure~\ref{fig:ligo_Daily_Byte} shows similar patterns as the daily numbers of data requests and cache misses in Figure~\ref{fig:dune_daily_count}.
For the DUNE project, the overall cache miss rate based on data volumes is 0.00024\% and the cache miss rate based on file requests is 0.00203\%. From the Table~\ref{tab:summary_data_Ams_projects}, the average file size for each cache miss is 2.4 MB, and the average file size for each file request is 20 MB. 
For the LIGO project, the overall cache miss rate in data volume is 0.56\%, and the cache miss rate for data access requests is 1.95\%. The average file size for each cache miss is 24 MB, and the average file size overall is 83 MB.


%% file: conc.tex
\section{Conclusions}
In this study, we explored the efficiency of cache utilization and the network traffic savings at the Cardiff and Amsterdam nodes from OSDF. 
Also, we studied data access and cache miss patterns for the DUNE and LIGO projects on the Amsterdam node.
Network traffic savings would be mostly from transatlantic data transfers.
Our study shows the potential for big improvements of backbone network performance from the impacts of the caching nodes.
From a total of 31,907,549 data accesses logged on both Cardiff and Amsterdam nodes from Oct. 2022 to Jan. 2023, we observed a total of 1.03 PB of client data accesses, with 2.5 TB of data transfers for cache misses from the remote data sources to the local cache, and 1.027 PB of network traffic volume savings from the repeated shared data accesses.
The Cardiff node reduced 96.95\% of the network traffic volume, and the Amsterdam node saved 99.86\% of the network traffic volume during the study period. 

With additional deployments of the caching nodes, we plan to study the access trends for a longer period of time and better understand the resource utilization which could lead to the efficient management of the data caches and long-term resource allocation planning.